\documentclass{aa}

\usepackage{graphicx,natbib}
\usepackage{amsmath}
\usepackage{ulem}

\bibpunct{(}{)}{;}{a}{}{,}

\def\ergscm{erg~s$^{-1}$~cm$^{-2}$}

\def\arcmin{\hbox{$^\prime$}}

\def\flux{erg s$^{-1}$ cm$^{-2}$}

\def\apj{ApJ}
\def\apss{Astroph.Sp.Sci.}
\def\aap{A\&A}
\def\mnras{MNRAS}

\begin{document}

\title{INTEGRAL constraints on the Galactic hard X-ray background from
  the Milky Way anticenter\thanks{Based on observations with INTEGRAL,
    an ESA project with instruments and science data centre funded by
    ESA member states (especially the PI countries: Denmark, France,
    Germany, Italy, Switzerland, Spain), Czech Republic and Poland,
    and with the participation of Russia and the USA}}

\author{R.~Krivonos\inst{1,2}, S.~Tsygankov\inst{1,2,3,4}, M.~Revnivtsev\inst{2}, S.~Sazonov\inst{2}, E.~Churazov\inst{1,2}, R.~Sunyaev\inst{1,2}}
\institute{
              Max-Planck-Institut f\"ur Astrophysik,
              Karl-Schwarzschild-Str. 1, D-85740 Garching bei M\"unchen,
              Germany
\and
              Space Research Institute, Russian Academy of Sciences,
              Profsoyuznaya 84/32, 117997 Moscow, Russia
\and
              Finnish Centre for Astronomy with ESO (FINCA), University of Turku,  V\"ais\"al\"antie 20, FI-21500 Piikki\"o, Finland
\and
Astronomy Division, Department of Physics, FI-90014 University of Oulu, Finland
            }
\authorrunning{Krivonos et al.}

\abstract{We present results of a study of the Galactic ridge X-ray
  emission (GRXE) in hard X-rays with the IBIS telescope on board
  INTEGRAL in the region near the Galactic Anticenter (GA) at
  $l=155^{\circ}$. We assumed a conservative $2\sigma$ upper limit on
  the flux from the GA in the $25-60$~keV energy band of
  $1.25\times10^{-10}$ \flux ($12.8$~mCrab) per IBIS field of view, or
  $6.6\times10^{-12}$ \flux ($0.7$~mCrab) per degree longitude in the
  $135^{\circ}<l<175^{\circ}$ region. This upper limit exceeds the
  expected GRXE intensity in the GA direction by an order of
  magnitude, given the near-infrared (NIR) surface brightness of the
  Milky Way in this region and the standard hard X-ray-to-NIR
  intensity ratio for the GRXE, assuming stellar origin. Based on the
  CGRO/EGRET surface brightness of the Galaxy above $100$~MeV as a
  tracer of the cosmic-ray (CR) induced gamma-ray background, the
  expected GRXE flux in GA exceeds the measured $2\sigma$ upper limit
  by a factor of~4. Therefore, the non-detection of hard X-ray
  emission from the GA does not contradict the stellar nature of the
  GRXE, but is inconsistent with CR origin.  }

\titlerunning{Constraints on the ridge emission from the Milky Way anticenter}
\maketitle

\section{Introduction}
\label{section:intro}

The stellar origin of the Galactic hard X-ray background, better known as the 
Galactic ridge X-ray emission (GRXE), has recently been strongly
supported by morphological/spectral studies with the RXTE and INTEGRAL
observatories \citep{mikej05,krietal07a,Turler10}, spectral studies
with Suzaku \citep{Ebisawa08,Yamauchi08} and direct X-ray source counts with
Chandra \citep{mikej09,mikej11}. The GRXE does not arise from the
interaction of cosmic rays with the interstellar medium, as was 
believed before, but is associated with the (predominantly old)
stellar population of the Galaxy, namely with hard X-ray
emission from accreting white dwarfs and coronaly active stars. It was
demonstrated \citep{mikej05,krietal07a} that the GRXE intensity closely follows
near-infrared (NIR) surface brightness over the Milky Way, which is a
known tracer of stellar mass. 

Galactic diffuse emission in a region far away from the Galactic
Center (GC) was studied by \cite{Worrall82} with the A2 experiment on
board the HEAO1 satellite. The observed $2-10$~keV emission was consistent
with an radial exponential disk with a half-thickness of $\sim240$ pc. It was
pointed out that unresolved emission likely comes from discrete point
sources and does not have a diffuse origin. In the hard X-ray domain
GRXE was studied by \cite{Skibo97} using OSSE observations at
longitude $l=95^\circ$. The observed emission between 50 and $600$~keV
was suspected to contain a significant contribution from bright
discrete sources because of the wide collimated field of view
($\sim11^{\circ}.4 \times 3^{\circ}.8$), but a major part of the
detected flux was interpreted as interstellar emission from
non-thermal electrons \citep[see also
  e.g.][]{Valinia2000,Valinia2001}. Today, thanks to the unique
possibilities of INTEGRAL gamma--ray telescopes, we can directly study
the Galactic diffuse background in any parts of the Galaxy without
dealing with significant source contamination.

Given the distribution of $4.9\mu$m NIR intensity over the Galaxy as
measured by the COBE/DIRBE experiment, the GRXE is not expected to be
detectable in the Galactic Anticenter (GA) because of the low NIR
surface brightness in this region. Nevertheless, an explicit
demonstration that the GRXE is not observed from a Galactic region of
low stellar density, such as the GA, would substantiate the stellar
paradigm of the GRXE even more. Placing tight constraints on the hard
X-ray flux from the GA region is also important for calibrating future
studies of the GRXE in the central parts of the Galaxy.

\section{Observations}
\label{section:data}
We used data from the ISGRI detector, the first layer of the IBIS
coded-mask telescope \citep{ibis}, on board the INTEGRAL gamma-ray
observatory \citep{integral}. ISGRI operates at energies above $\sim
20$~keV, with the sensitivity rapidly decreasing above $100$~keV. IBIS
has a relatively wide field of view ($\sim 28^\circ \times 28^\circ$
at zero response), which allows one to measure weak diffuse emission
fluxes by using the telescope as a collimated instrument.

To study the GRXE in the GA region, we used special INTEGRAL
observations, part of a series of so-called Galactic latitude scans
(\textit{GLS}). This program is based on consecutive observations made
along the Galactic latitude in the range $\pm30^\circ$ with a step of
$2^\circ$. This strategy allows one to make independent snapshot
measurements of the instrumental background at mid latitudes, where
the GRXE is expected to be negligible, along with an actual GRXE
observation near the Galactic plane. It is crucial that the
instrumental background, which is usually high in hard X-ray
observations, does not exhibit strong variability during an individual
\textit{GLS} lasting $\sim8$~hours.

There are two positions in the GA region observed with \textit{GLS}s:
ongoing observations at $l=215^\circ$ (PI: Tsygankov) and a completed
program at $l=155^\circ$ with a total exposure of $1$~Ms (PI:
Krivonos). In the present study we only used the completed
\textit{GLS} observations at $l=155^\circ$ performed in
August--September 2010, see Table~\ref{tab:observations} for
details. After screening the whole data set following the procedure
described in \cite{krietal07a}, hereafter K07, we were left with 525
out of 569 short ($\sim2$~ks) individual INTEGRAL observations, called
\textit{scientific window}s (\textit{ScW}s), for the subsequent
scientific analysis. To model background variations we used public
data of high-latitude observations (Table~\ref{tab:observations}).

\begin{table}[h]
\caption{INTEGRAL observations used for the GRXE study.}
\label{tab:observations}
\centering         
\begin{tabular}{l l l}  
\hline                                   
\multicolumn{3}{c}{Latitude scan at $l=155^\circ$, 2010}\\
\hline\hline                
Observation field & Orbits & \textit{ScW}s \\   
\hline                                   
... & 960,961 &2--94,11--97 \\
... & 962,963 &12--112,1--97 \\
... & 964,965 &13--110,2--67 \\
... & 966 &23--56 \\
\hline                                   
\multicolumn{3}{r}{569 \textit{ScW}s, total nominal exposure: 1~Ms.}\\
\hline                                   
\hline
\multicolumn{3}{c}{Background model calibration, 2008--2009}\\
\hline
\hline
Observation field & Orbits &  \\   
\hline
North Ecliptic Pole & \multicolumn{2}{l}{686--689, 759--761, 824--829}    \\
XMM-LSS &695, 696, 701 & \\
Virgo Cluster & \multicolumn{2}{l}{747--754, 758, 819--820}   \\
Coma Cluster &821--823 &  \\
M82 X-1 &869--872 &  \\
3C273, 3C279 and M87 &878--880 &  \\
\hline                                   
\multicolumn{3}{r}{Total nominal exposure: 6~Ms.}\\
\hline                                   
\hline
\multicolumn{3}{c}{Crab calibration, 2010}\\
\hline
\hline
Observation field & Orbits &  \\   
\hline                                   
Crab Nebula & \multicolumn{2}{l}{902, 903, 966-968, 970} \\
\hline
\multicolumn{3}{r}{Total nominal exposure: 890~ks.}\\
\hline                                   
\end{tabular}
\end{table}

\section{Analysis}
\label{section:analysis}
We mainly followed the approach described in K07. 
Our study of the GRXE is based on the capability of the IBIS telescope
to separate the contributions of point sources from the background
count rate. 

The ISGRI detector shadowgram that was accumulated during individual
INTEGRAL observation in a given energy range was cleaned from source
fluxes using the known pattern of the IBIS mask. The remaining
shadowgram contains the following components: (i) variable
instrumental background, (ii) the isotropic cosmic X-ray background
(CXB), and (iii) the GRXE. Throughout the analysis we assumed the CXB
flux as a constant part of the instrumental background. The possible
influence of the CXB variance on the GRXE measurement is considered in
Appendix~\ref{section:cxb}. Because the GRXE is extended over the sky,
it cannot be directly resolved with the IBIS mask. The only way to
estimate the GRXE flux is to determine the difference between the
observed collimated detector count rate that is cleaned from sources
and the assumed instrumental background. To this end, we define a
background model in Appendix~\ref{section:model} that predicts the
background count rate during a given GRXE observation. The background
model is adjusted very precisely using mid-latitude snapshots of the
background performed shortly before and after a given GRXE
observation, which is the main concept of the \textit{GLS}.

\subsection{Detector shadowgram}
\label{section:detector}
Because IBIS is a coded aperture imaging telescope, the sky is projected
onto the detector plane through the transparent and opaque elements of
the mask mounted above the detector plane. We produced the ISGRI detector
shadowgram for every \textit{ScW} as described in K07. We used the
$25-60$~keV working energy band because of the known evolution of the 
low-energy threshold of the ISGRI detector and because the GRXE 
is expected to be weak above $\sim50$~keV owing to the high-energy
cut-off in the GRXE spectrum \citep{krietal07a,mikej05}.

\subsection{Sky map}
\label{section:sources}

\begin{figure}[t]
 \includegraphics[width=0.48\textwidth]{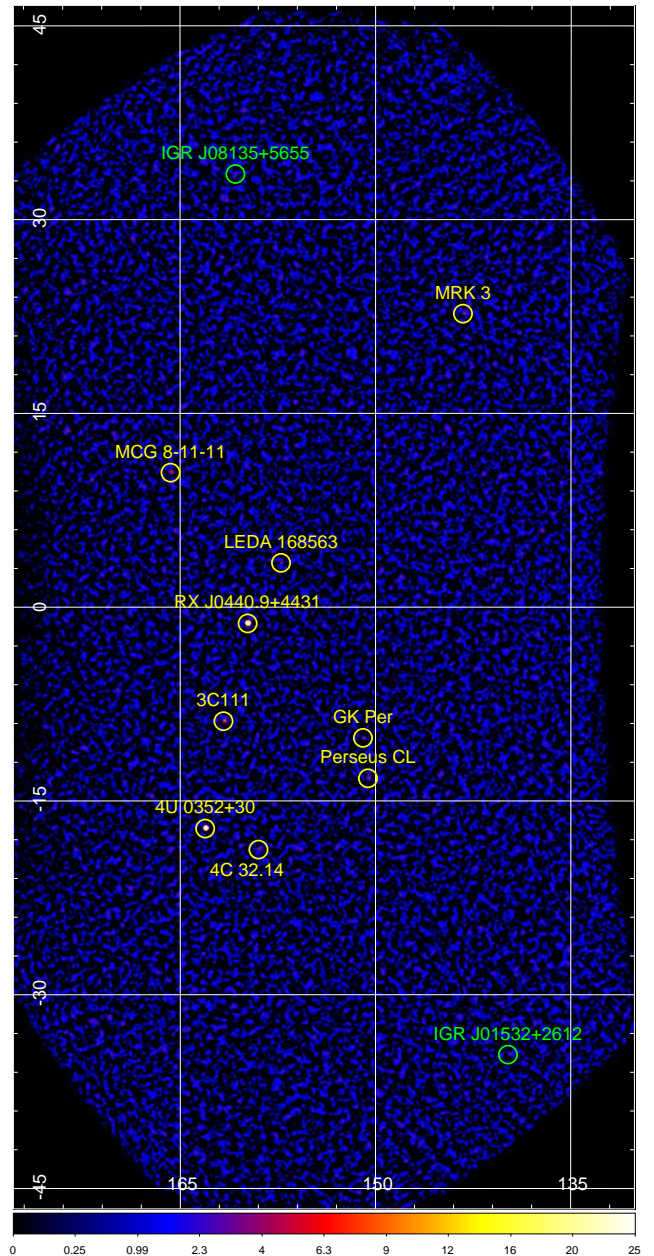}
\caption{IBIS/ISGRI sky image of the $1$~Ms observation of the GA
  region at $l=155^{\circ}$ produced in the $25-60$~keV energy
  range. The color table of the image represents pixel values in the
  range 0-25 calculated as the square root of the significance. The
  cataloged and newly detected sources are labeled in yellow and
  green, respectively.  }\label{fig:map}
\end{figure}

The sky reconstruction is based on deconvolution of the detector
shadowgram with a known mask pattern. We implemented the IBIS/ISGRI sky
reconstruction method described in our previous publications
\citep{revetal04,krietal05,krietal07a,krietal07b}. For the basic idea
we refer the reader to the papers by \cite{fenimore81} and
\cite{skinner87}. Every sky image was additionally cleaned from
systematic noise as described in \cite{krietal10a}. The resulting sky image
mosaic is shown in Fig.~\ref{fig:map}.

\begin{figure}[h]
  \includegraphics[width=0.47\textwidth,bb=18 156 570 520]{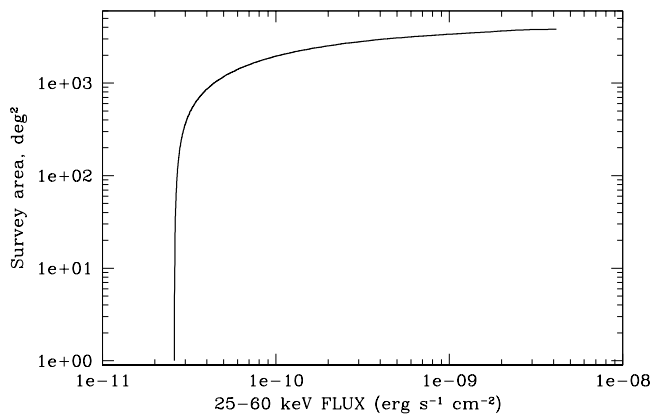}
  \caption{Covered area of the survey as a function of flux for sources with $S/N>5$}\label{fig:area}
\end{figure}

\begin{figure}[h]
 \includegraphics[width=0.47\textwidth]{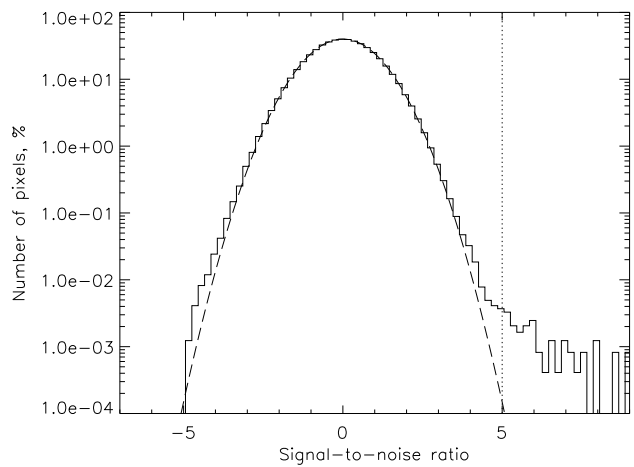}
\caption{Signal-to-noise ratio distribution of pixels in
  the 1~Ms \textit{GLS} observation at $l=155^\circ$. The dashed line
  represents the normal distribution with unit variance and zero
  mean. The accepted threshold of source detection ($5\sigma$) is
  shown by the dotted line. The plot is truncated at $\sigma=9$ for
  convenience.}\label{fig:snr}
\end{figure}

The survey area as a function of flux for sources with $S/N>5$ is
shown in Fig.~\ref{fig:area}. The minimum detectable flux in the
central part of the field is $2.4\times10^{-11}$ \flux (or
$2.5$~mCrab). The survey area reaches its geometric limit of $3650$
deg$^2$ for $f>4.9\times10^{-9}$\flux ($500$~mCrab), $50\%$ of this area
has a sensitivity better than $8.1\times10^{-11}$\flux ($8.3$~mCrab).

We performed a search for sources as excesses in the sky mosaic
(Fig.~\ref{fig:map}) convolved with a Gaussian representing the
effective instrumental PSF. The detection threshold was estimated
assuming Gaussian noise of the pixel values as follows. The total area
of the sky image (Fig.~\ref{fig:map}) is 3650 squared degrees, and
taking the IBIS telescope angular resolution of $12$\arcmin\ into
account, we gathered $\sim9.12\times 10^{4}$ independent pixels. With
this consideration, we set the source detection threshold to
$5\sigma$, allowing at most one false detection \textit{for pure
  Gaussian noise}.

%print,erfc(4.4/sqrt(2))*91250

%Because of good quality of the sky mosaic in Fig.~\ref{fig:map}, 

The signal-to-noise ratio distribution of pixels shown in
Fig.~\ref{fig:snr} has the expected Gaussian shape. However, one
notices some systematic excess at the negative side. Nevertheless, the
source detection threshold of $5\sigma$, estimated above,
separates the noise and source dominated pixel domains. 

The list of point sources is presented in Table~\ref{tab:sources}. We
attribute two marginally detected, previously unknown sources to the
systematic noise. IGR~J08135+5655 is not detected in the sky mosaic of
the slightly broader energy band $20-60$~keV, and the region around
IGR~J01532+2612 is affected by systematic noise at the edge of the sky
mosaic.

\begin{figure*}[t]
  \includegraphics[width=1.0\textwidth,bb=29 185 570 340]{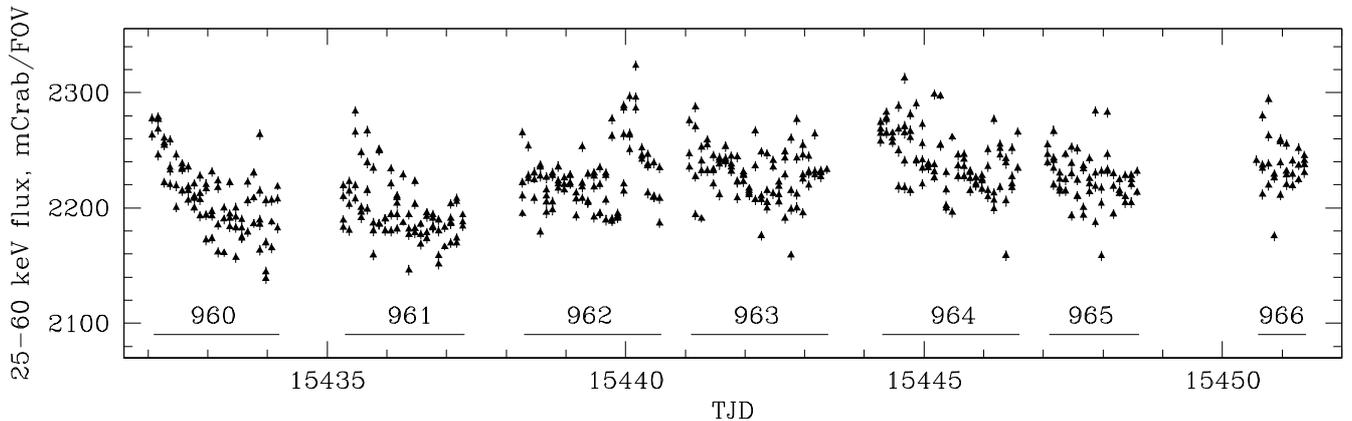}
  \caption{Detector count rate of the individual \textit{ScW}s after
    subtracting flux contribution from point sources during the \textit{GLS} program
    at $l=155^{\circ}$. The data points contain pure statistical
    errors. The labeled horizontal intervals denote different
    spacecraft orbits.}\label{fig:data}
\end{figure*}

The list of sources in Table~\ref{tab:sources}, except for two IGR's
mentioned above, was used in the iterative source removal (IROS)
procedure that we applied to every ISGRI detector shadowgram
\citep[e.g.][]{krietal10a}. This procedure introduces additional
uncertainty to the background model (Appendix~\ref{section:accuracy}),
but allows one to trace source variability. For example, the known
Be/X-ray binary RX~J0440.9+4431 (LS~V~+44~17) was in a strong outburst
during the observations \citep{atel2828,sst11}. Fig.~\ref{fig:data}
shows $25-60$~keV detector light curve, cleaned from the source
contribution, and ready for further analysis.

%
% S=5431.38 sq. degrees
% angular resolution A=(12/60)**2
% number of independent pixels  S/A=135784

\begin{table}[t]
\caption{The list of sources significantly ($\ge 5\sigma$) detected
  on the sky mosaic (Fig.~\ref{fig:map}). The newly detected sources are
  highlighted in bold. The $68\%$ confidence interval for estimated
  sky coordinates depends on source significance: $2.1$\arcmin,
  $1.5$\arcmin, and $<0.8$\arcmin for $5-6$,10, and $>20\sigma$,
  respectively \citep{krietal07b}}
\label{tab:sources}
\centering         
\begin{tabular}{l r r r r r}  
\hline\hline                
Name & lon. & lat. & $F_{\rm 25-60~keV}$  \\   
 & deg. &  deg. & mCrab \\   
\hline
4U 0352+30$^{\mathrm{a}}$ & 163.08 & -17.14 & 39.42 $\pm$ 0.69 \\% & 57.33 \\
RX J0440.9+4431$^{\mathrm{a}}$ & 159.82 & -1.26 & 26.12 $\pm$ 0.53 \\% & 48.82 \\
3C111$^{\mathrm{b}}$ & 161.68 & -8.83 & 5.32 $\pm$ 0.60 \\% & 8.84  \\
MCG 8-11-11$^{\mathrm{b}}$ & 165.74 & 10.41 & 7.27 $\pm$ 0.86 \\% & 8.41 \\
4C 32.14$^{\mathrm{c}}$ & 158.99 & -18.77 & 3.15 $\pm$ 0.56 \\% & 5.63 \\
LEDA 168563$^{\mathrm{b}}$ & 157.26 & 3.43 & 2.76 $\pm$ 0.49 \\% & 5.58 \\
MRK 3$^{\mathrm{b}}$ & 143.29 & 22.72 & 7.23 $\pm$ 1.33 \\% & 5.43  \\
Perseus CL$^{\mathrm{d}}$ & 150.58 & -13.25 & 3.06 $\pm$ 0.58 \\% & 5.30 \\
\textbf{IGR J01532+2612} & 139.82 & -34.64 & 20.31 $\pm$ 3.91 \\% & 5.19 \\
\textbf{IGR J08135+5655} & 160.75 & 33.53 & 7.09 $\pm$ 1.38 \\% & 5.14 \\
GK Per$^{\mathrm{e}}$ & 150.96 & -10.12 & 2.82 $\pm$ 0.56 \\% & 5.07 \\
%\textbf{IGR J07274+5531} & 161.81 & 27.04 & 4.22 $\pm$ 0.87 & 4.85 & &\\
\hline                                   
\end{tabular}
\begin{list}{}{}
\item[$^{\mathrm{a}}$] HMXB, $^{\mathrm{b}}$ AGN, $^{\mathrm{c}}$ QSO, $^{\mathrm{d}}$ Cluster of Galaxies, $^{\mathrm{e}}$ CV
\end{list}

\end{table}

% new sources:
%IGR J08135+5655         , 123.38 , 56.92 , 695
%IGR J01532+2612         , 28.31 , 26.20 , 696
%IGR J03129+6444         , 48.24 , 64.73 , 697
%IGR J07274+5531         , 111.85 , 55.53 , 698

\section{Results}
\label{section:results}
\begin{figure*}[t]
  \includegraphics[width=1.0\textwidth,bb=29 185 570 340]{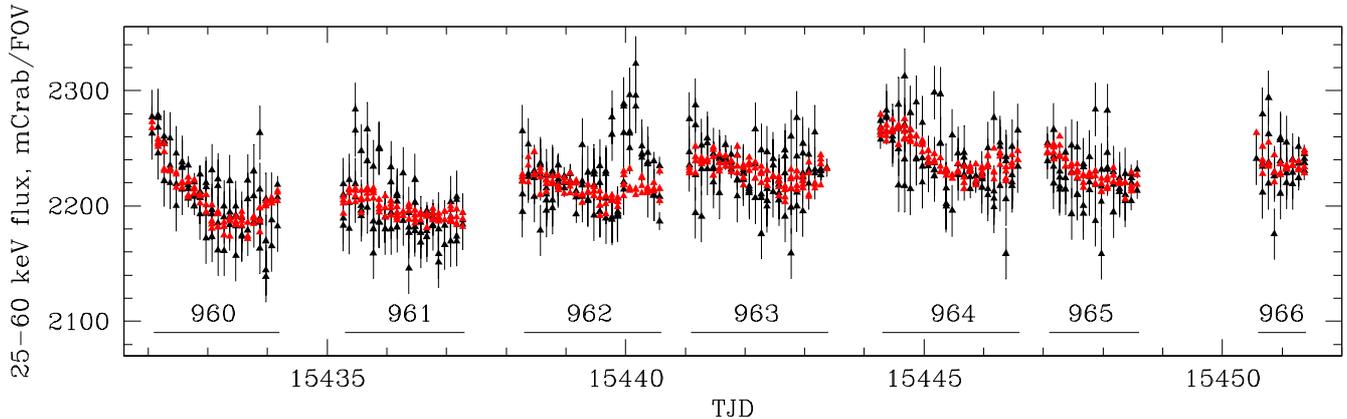}
  \caption{Detector count rate (black) of the individual \textit{ScW}s
    as shown in Fig.~\ref{fig:data}. The systematic uncertainty of the
    source removal procedure (Appendix~\ref{section:accuracy}) was added to the
    statistical errors of each point. The red points represent the
    count rate predicted by the background model.}\label{fig:datafit}
\end{figure*}

\begin{figure}[t]
  \includegraphics[width=0.49\textwidth,bb=18 156 570 520]{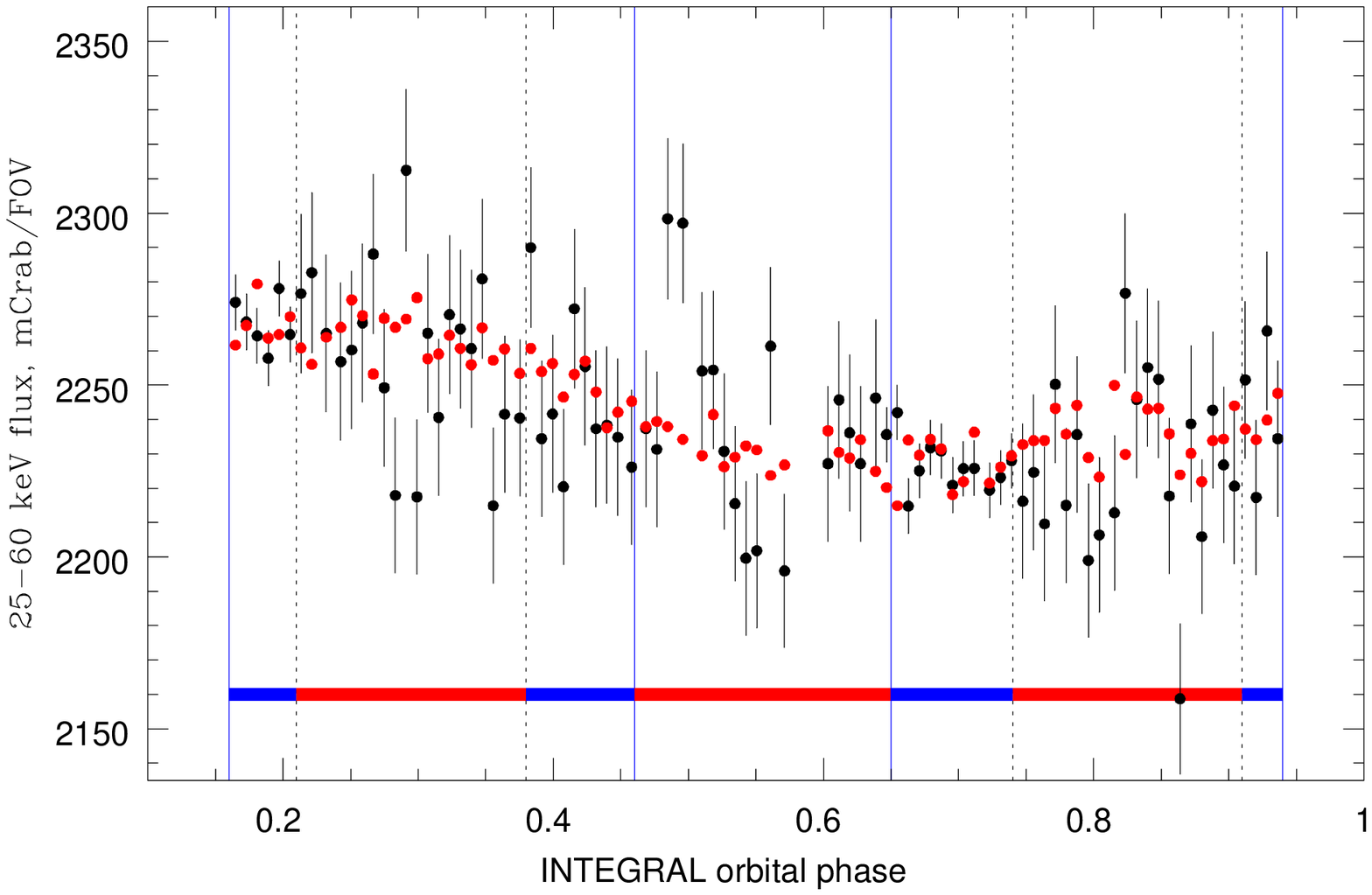}
  \includegraphics[width=0.49\textwidth,bb=18 156 570 520]{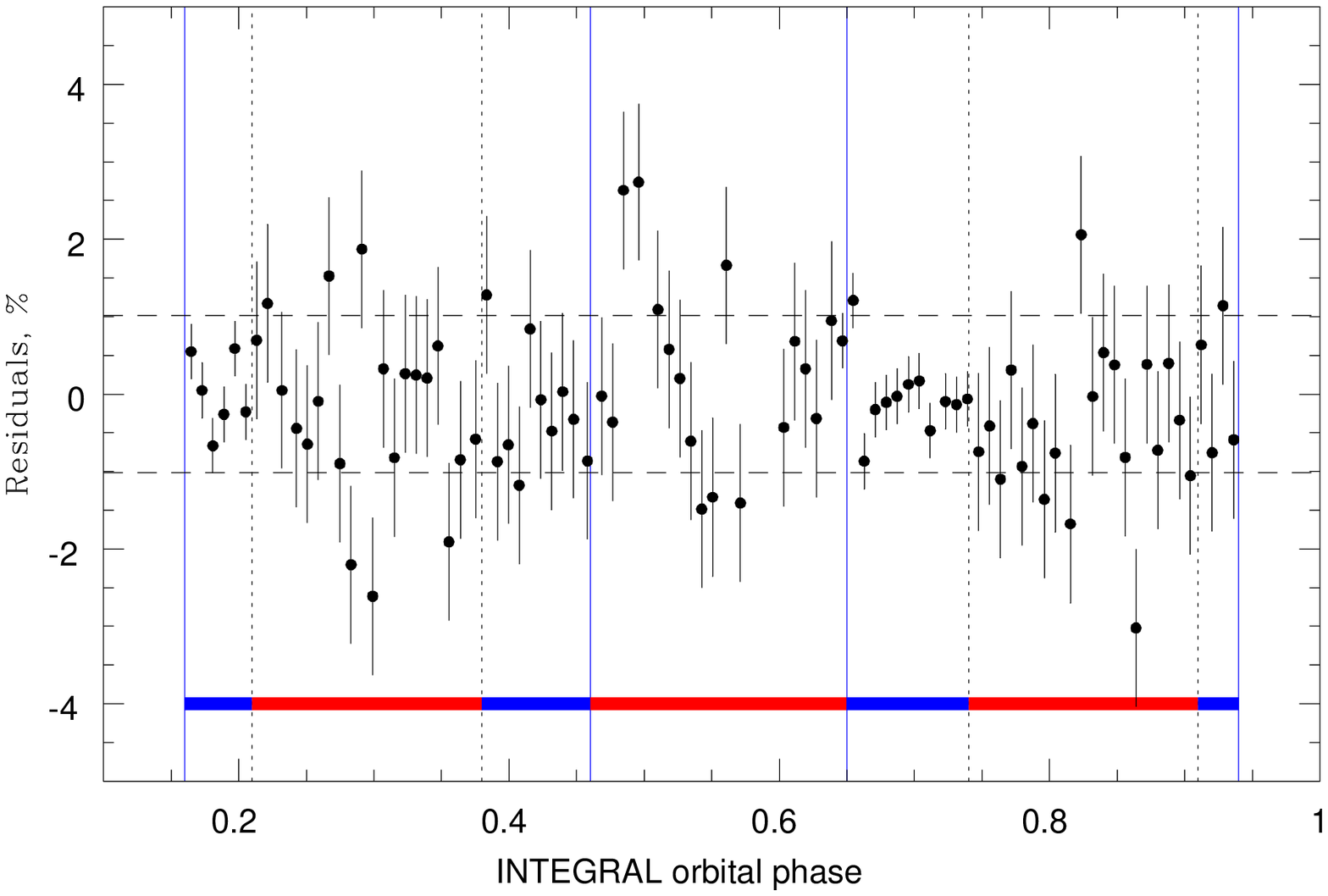}
  \caption{\textit{Upper panel:} Detailed view of detector count
    rate (black) during orbit 964 from Fig.~\ref{fig:datafit}. The
    blue and red areas denote observations made at $|b|>=20^{\circ}$
    and $|b|<20^{\circ}$, respectively. \textit{Bottom panel:}
    Residuals after subtracting the model-predicted count rate from
    the observed count rate. The black dashed lines represent a
    $1\sigma$ deviation ($1.02\%$) from zero.}\label{fig:fit}
\end{figure}

Using the background model (Appendix~\ref{section:model}) we obtained
for each \textit{ScW} the predicted detector count rate, as shown in
Fig.~\ref{fig:datafit}. The detailed view of the observed and
predicted detector light curves and their residuals during a
spacecraft orbit 964 is shown in Fig.~\ref{fig:fit}. We denoted
regions of background measurement ($|b|>=20^{\circ}$) in blue, and
actual GRXE observations ($|b|<20^{\circ}$) in red. Fig.~\ref{fig:fit}
shows that the background behavior in an individual orbit can be
captured only with fast scanning observations such as
\textit{GLS}s. The background model, with low intrinsic scatter,
exactly follows the observed detector rate, and the scatter of
residuals (lower panel of Fig.~\ref{fig:fit}) is comparable to the
statistical uncertainty of the IROS procedure
(Appendix~\ref{section:accuracy}).

\begin{figure}[t]
 \includegraphics[width=0.5\textwidth]{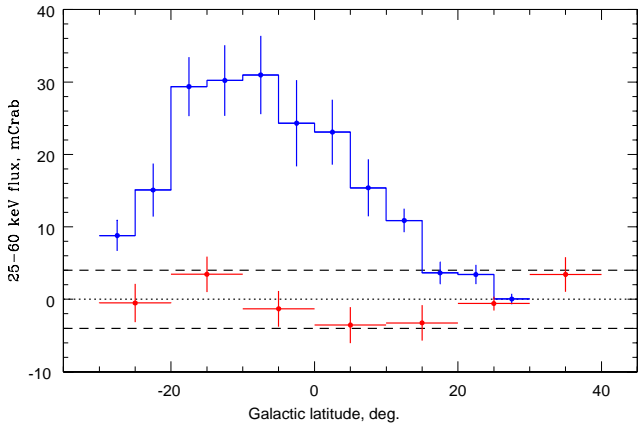}
\caption{Source flux contribution (blue) and ISGRI detector background
  count rate residuals (red) averaged over the Galactic
  latitude. Error bars of the blue histogram represent rms-deviations
  of the summed point source fluxes from average in
  bin. }\label{fig:ridge}
\end{figure}

Using the entire \textit{GLS} data set at $l=155^{\circ}$, we averaged
residuals over Galactic latitude, as shown in Fig.~\ref{fig:ridge} by
red points. The latitude profile does not show any significant
excess in the Galactic plane region at $b=0^\circ$. As expected, the
GRXE associated with the old stellar population is not detected in the
GA. The $1\sigma$ upper limit on the GRXE flux in the $|b|<5^\circ$
latitude range, which roughly corresponds to the IBIS fully coded FOV,
is $\sim 2.8$~mCrab, or $\sim 6.4$~mCrab taking systematic
uncertainties into account. We later refer to a $2\sigma$ upper limit
of $\sim 12.8$~mCrab. One can convert the achieved upper limit to
more convenient units using the effective solid angle of the IBIS
telescope $\sim286$ deg$^2$ and taking into account that the GRXE is
much less extended in the Galactic latitude than the average
cross-section of IBIS FOV. For instance, the $2\sigma$ upper limit on
the GRXE flux at $l=155^{\circ}$ per unit Galactic longitude is
$0.7$~mCrab~deg$^{-1}$.
%or $6.56\times10^{-12}$ \flux ~deg$^{-1}$.}

\begin{figure}[t]
 \includegraphics[width=0.5\textwidth,bb=18 156 570 520]{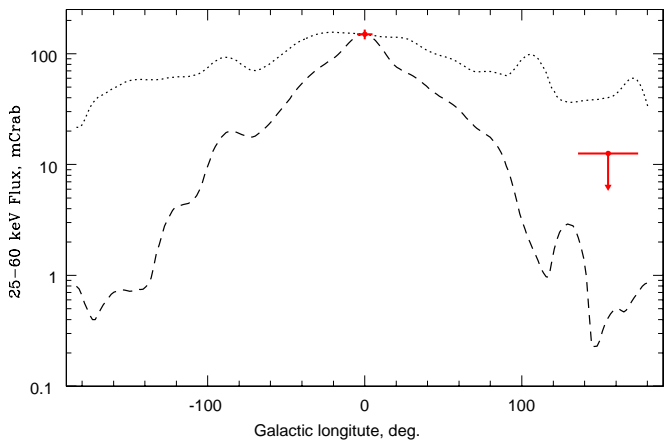}
\caption{Galactic longitude profiles of the COBE/DIRBE $4.9\mu$m intensity
  (dashed line) and EGRET background above $100$~MeV (dotted line),
  both normalized to the GRXE flux of $150$ mCrab (red point) in the
  GC. The $2\sigma$ upper limit corresponds to the GRXE measurement at 
  $l=155^{\circ}$ in the present study.}\label{fig:galprof}
\end{figure}

One can test the GRXE non-detection in the GA for consistency with
stellar and truly diffuse GRXE origins. To this end, we compared the
observed drop of the hard X-ray flux from the GC to the GA region (per
IBIS FOV) with the corresponding change of the intensity of a 
given tracer. We used the COBE/DIRBE $4.9\mu$m data\footnote{COBE/DIRBE
  $4.9\mu$m intensity map was corrected for the interstellar reddening
  and mean background level measured in high-latitude regions, as
  described in K07.} as a tracer of stellar mass, and the EGRET
gamma-ray background map above $100$~MeV as a tracer of the cosmic-ray
induced gamma-ray background. The EGRET background intensity drops
from the GC to the GA by a factor of $\sim3$. Therefore, using the
conservative estimate of the GRXE flux in the GC from K07 of
$150\pm15$~mCrab, we expect the corresponding flux from the GA to be
$\sim50$~mCrab. This is definitely not observed according to
Fig.~\ref{fig:ridge}. In contrast, there is a factor of $270$ drop in
the NIR $4.9\mu$m intensity from $2.7 \times 10^{-5}$ \ergscm\ to $10^{-7}$ 
\ergscm. This implies a GRXE flux from the GA of $0.4$~mCrab at
$25-60$~keV, which is consistent with the derived upper limit of 
$12.8$~mCrab. This is illustrated in
Fig.~\ref{fig:galprof}, where the COBE/DIRBE and EGRET longitude profiles
are renormalized to the hard X-ray flux observed from the GC. We conclude
that the non-detection of the GRXE from the GA is consistent with the stellar 
mass distribution of the Galaxy traced by NIR maps, rather than with
the cosmic-ray induced gamma-ray background seen by EGRET.

%We note, however, that the NIR flux measured in the GA region is
%expected to be dominated by relatively nearby (with respect to the
%Sun) stars, therefore the NIR intensity map measured by COBE/DIRBE may
%not be an accurate tracer of stellar mass in the GA. On the other
%side, the existing models of the Galactic stellar density distribution
%in the disk have large uncertainty in the normalization, and fail to
%describe the irregular outskirts of the Milky Way.

\section{Conclusions}

1) Using the $1$~Ms observations of the GA at $l=155^\circ$ with the 
INTEGRAL observatory, performed in the special \textit{GLS} mode, we
did not detect the GRXE in the $25-60$~keV energy band and set a
conservative $2\sigma$ upper limit of $1.25\times10^{-10}$ 
\flux ($12.8$~mCrab) per IBIS FOV or $6.6\times10^{-12}$ \flux
~deg$^{-1}$ ($0.7$~mCrab~deg$^{-1}$) per unit Galactic longitude.

2) The obtained upper limit is consistent with the considerable drop in the
NIR ($4.9\mu$) intensity observed by COBE/DIRBE and disagrees with the
much smaller decrease in the gamma-ray (above $100$~MeV) background
measured by EGRET. Therefore, the non-detection of the GRXE in the GA
is consistent with the stellar mass distribution in the Galaxy, which
does not contradict the stellar nature of GRXE, but is inconsistent with
its CR origin.

3) The developed background model potentially allows one to reach the
statistically limited accuracy. However, the final uncertainty of the
approach is associated with the source removal procedure, the
systematic uncertainty of the method itself, and the CXB
variance. Nevertheless, the implemented method along with the special
\textit{GLS} mode of observation is an optimal approach of modeling
the ISGRI background and can be efficiently used for studying the
Galactic hard X-ray background.

\begin{acknowledgements}
This research was made possible thanks to the unique capabilities of the
INTEGRAL observatory. The data used were obtained from the European
and Russian INTEGRAL Science Data Centers. The work was supported by
the President of the Russian Federation (through the program of
support of leading scientific schools, project NSH-5069.2010.2, and
grant MD-1832.2011.2), by the Presidium of the Russian Academy of
Sciences/RAS (the program ``Origin, Structure, and Evolution of
Objects of the Universe''), by the Division of Physical Sciences of
the RAS (the program ``Extended objects in the Universe'', OFN-16), by
the Russian Basic Research Foundation (grant 10-02-00492-A), State
contract 14.740.11.0611, and the Academy of Finland grant 127512.
\end{acknowledgements}

\begin{appendix}

\section{Background model}
\label{section:model}

The background model used in K07 was slightly changed in this
study. Since the \textit{GLS} observations were performed over a
relatively short time period, we removed the long-term time part from
the equation. Instead of using the gain parameter to trace orbital
modulations of the background rate, we used the spacecraft orbital
phase $P$ in a quadratic polynomial form. Hereafter, we consider only
the detector count rates after removal of the contribution of point
sources. The model of the detector background $25-60$~keV count rate,
$D_{bgd}$, is made of a linear combination of the $600-1000$~keV
detector count rate, $H$, and phase

\begin{equation}
%\begin{split}
D_{bgd}=const + a H + b_1 P + b_2 P^2.
%\end{split}
\label{eq:model}
\end{equation}
The coefficient $a$ was calculated using observations pointed away
from the Galactic plane ($|b|>20^\circ$) where the GRXE is not
expected to be observed (see Table~\ref{tab:observations}). The
constant term and $b$ coefficients were determined individually for
each spacecraft orbit from the observations at $|b|>20^\circ$, thus
adjusting the model to the current background level. In fact, the
constant term in Eq.~\ref{eq:model} contains contribution from CXB and
unknown intrinsic detector background. The last is also variable,
which is traced by variability of this constant with time. In the
current observations, its absolute value varies in the range of
$800-900$~mCrab from orbit to orbit, while permanent CXB contribution
is expected to be at level of $550$~mCrab
(Appendix~\ref{section:cxb}). Finding the difference between the
observed and predicted by Eq.~\ref{eq:model} detector count rate
should yield the GRXE excess in the Galactic plane.

The detector count rate was converted to the convenient units of Crab
flux, with $1$~mCrab~$=7.25\times 10^{-6}$~cts/s in the $25-60$~keV
band per IBIS FOV. The conversion coefficient was determined from 
observations of the Crab Nebula in 2010
(Table~\ref{tab:observations}). A flux of $1$~mCrab in the $25-60$~keV 
energy band corresponds to $9.7\times10^{-12}$ \flux\ for a source
with a Crab-like spectrum, $10.0\times
E_{\text{keV}}^{-2.1}$~phot~cm$^{-2}$~s$^{-1}$~keV$^{-1}$.

\section{CXB cosmic variance}
\label{section:cxb}

%Throughout the analysis we assumed the CXB flux to be constant and
%isotropic. However, the 

The CXB emission coming from the population of unresolved extragalactic
sources (active galactic nuclei, AGNs) is subject to Poissonian
variations in the number of sources, intrinsic source variability, and
nearby large-scale structure \citep[see e.g.][]{Fabian92}. Here, we
estimate the systematic limitations to the measured GRXE flux caused
by CXB variations.

%% \begin{figure}[t]
%%   \includegraphics[width=0.5\textwidth,bb=18 156 570 520]{Maps/lognlogs.ps}
%%   \caption{Number--flux relation for non-blazar extragalactic objects
%%     detected in the survey. The dotted line represents the $\log N$--$\log
%%     S$ relation from \cite{krietal10b}.}\label{fig:lognlogs}
%% \end{figure}

%% During the GA survey we detected an enhanced number of extragalactic
%% sources: four nearby AGNs, a QSO and a cluster of galaxies
%% (see Table~\ref{tab:sources}). Fig.~\ref{fig:lognlogs} shows the
%% number-flux relation of the detected AGNs in comparison with the best-fit
%% $\log N$--$\log S$ function of 158 non-blazar AGNs derived over the
%% extragalactic sky ($b>5^{\circ}$) in the 7-year INTEGRAL survey
%% \citep{krietal10b}. This $\log N$--$\log S$ relation was converted
%% from $17-60$~keV to $25-60$~keV assuming a power-law spectrum with a
%% slope of $1.8$.

Using the extragalactic $\log N$--$\log S$ relation from
\cite{krietal10b} and following \cite{mikej08}, we estimated the 
relative uncertainty of the CXB flux in the $25-60$~keV band as

\begin{equation}
\left(\frac{\delta I_{CXB}}{I_{CXB}} \right)_{\Omega}\sim 5.5\times 10^{-2} {S_{\rm max,11}^{1/4} \Omega_{\rm deg} ^{-1/2}},
\label{eq:poiss}
\end{equation}
where $S_{\rm max,11}$ is the maximum flux of undetected sources in
units of $10^{-11}$ \flux, and $\Omega_{\rm deg}\approx286$ is the
effective solid angle of the IBIS telescope. We adopted the CXB
intensity to be equal to $1.89\times10^{-11}$ \flux\ deg$^{-2}$, based
on the CXB spectrum model of \cite{gruber99} and the $\sim10\%$ higher
normalization measured by INTEGRAL \citep{churazov07}. Using the
limiting flux of the survey Eq.~\ref{eq:poiss} yields a CXB variance
at the level of $\sim0.4\%$. The absolute value is $2.3$~mCrab
assuming a $\sim550$~mCrab CXB flux per IBIS FOV. In the current work,
we consider the CXB variance for an area of three IBIS FOVs, which
approximately corresponds to the effective area of the GA survey.

\section{Uncertainty in GRXE measurements}
\label{section:accuracy}

The uncertainty of the background model, i.e. the accuracy of the ISGRI
background rate prediction, is subject to statistical and systematical
errors. The former can be easily estimated from the total number of
counts ($\sim 3\times 10^5$) in the $25-60$~keV energy band per
typical \textit{ScW} ($\sim2$~ks). 

The additional statistical effect is related to the IROS procedure,
when the count rate attributed to a given source is removed from the
detector using the known aperture function of the mask. To a first
approximation, the total number of counts, $S$, associated with a given
source is determined as the difference between the number of counts,
$D_1$, in the detector pixels illuminated by the source through the 
mask and the number of counts, $D_0$, in the detector pixels blocked by
the mask: $S=D_1-D_0$. The total flux on the detector is
$D=D_0+D_1$. Thus, subtracting the contribution of the source yields
the residual detector flux $D^{\prime}=D-S=2D_0$. For a weak source in
the center of the field of view, $D_0 \approx D_1 \approx 1/2\times D$
and therefore the relative statistical uncertainty of measuring the
detector count rate
$\frac{D^{\prime}}{\sqrt{2D_0}}=\sqrt{2}\frac{D}{\sqrt{D}}$ increases
by a factor of $\sqrt{2}$. In practice, a more complicated model of a
point source (see K07), implemented in the source subtraction
algorithm, causes an even larger increase of statistical
uncertainties. 

\begin{figure}
  \includegraphics[width=0.5\textwidth]{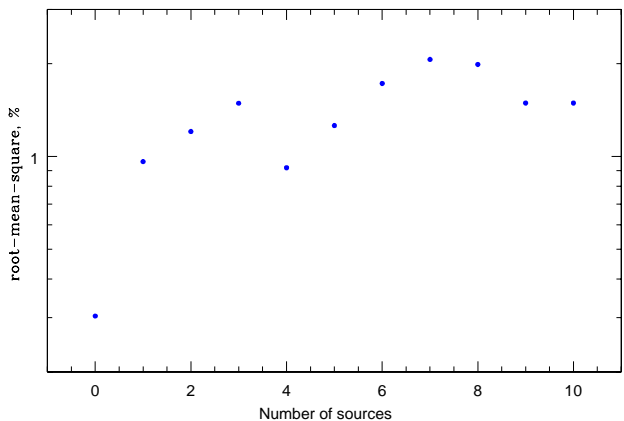}
  \caption{Relative root-mean-squared value of the detector rate
    as a function of number of sources in the FOV.}\label{fig:scatter}
\end{figure}

To estimate the IROS induced uncertainty as a function of number of
sources in the FOV, we studied a set of 22 consecutive observations
without cataloged and detected sources (orbit 973, \textit{ScW}s
$58-80$). The relative standard deviation of detector count rate, as a
function of the number of simulated sources, $N_{src}$, is shown in
Fig~\ref{fig:scatter}. The first point at $N_{src}=0$, $RMS=0.3\%$,
reflects the relative statistical scatter of the data. As seen from
Fig~\ref{fig:scatter}, the scatter rapidly increases with inclusion of
sources in the FOV. A typical scatter of $\sim1.0\%$ on the
\textit{ScW} time scale is observed in real data, as demonstrated in
Sect.~\ref{section:results}.

\begin{figure}[t]
  \includegraphics[width=0.5\textwidth]{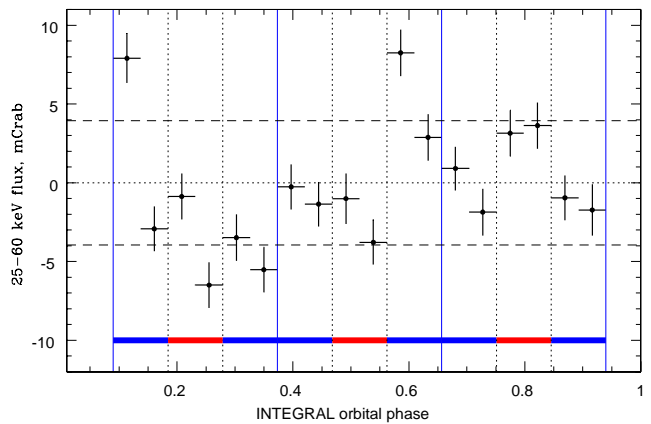}
  \caption{Residuals after subtracting the model-predicted count rate
    from the observed count rate for the $5.3$Ms high-latitude
    observations averaged over spacecraft orbital phase. The black
    dashed lines represent a $1\sigma$ deviation ($4.0$~mCrab) of the
    averaged values from zero. Blue and red regions denote different
    phase intervals used for background model calibration and actual
    measurements, respectively.}\label{fig:uncert}
\end{figure}

To estimate the systematic uncertainty of the method, we defined
coefficients of the background model in Eq.~\ref{eq:model} using the
high-latitude observations (see Table~\ref{tab:observations}) in the
South Galactic hemisphere (700~ks) and applied it to the North
(5.3~Ms). The non-existent GRXE flux was averaged over a given INTEGRAL
orbit divided into the three equal intervals having three parts: one
in the middle and two adjacent, see Fig.~\ref{fig:uncert} for 
reference. The middle part of each interval (in red) was supposed to 
have GRXE flux, and the neighboring parts (in blue) were used to correct the
constant term (Eq.\ref{eq:model}). This set-up mimics the \textit{GLS}
pattern of observations.

The standard deviation of residuals from zero represents the
systematic uncertainty of our background model, which is found to be $\sim
4.0$~mCrab. The $25-60$~keV detector count rate increased from $\sim2.2$
to $\sim2.5$ Crab over the considered time period, hence the relative
accuracy of the model is $\sim0.17\%$ of the observed background rate. 

We summarize all the discussed uncertainties related to the GRXE
measurements in the $25-60$~keV energy band. The values below are
presented with respect to the background rate, which is assumed to be
$2.5$~Crab. \textit{Statistical uncertainties:}
\begin{itemize}
\item 0.18$\%$ ($4.5$~mCrab) -- count statistics, \textit{expected}
  for $3\times10^5$ counts per typical ScW ($\sim2$~ks), 
\item 0.30$\%$ ($7.5$~mCrab) -- count statistics, \textit{observed} in
  a typical ScW without any sources in the FOV. Since this value
    differs from the expected, it cannot be fully statistical. Some
    unexplored systematics or background variability can contribute to
    the scatter of the observed detector count rate.
\item 1.00$\%$ ($25$~mCrab) -- observed in a typical ScW with several
  sources in the FOV, related to the IROS procedure.
\end{itemize}
These uncertainties constitute the error of measurement and,
naturally, decrease with increasing exposure. For example, the largest
error of $25$~mCrab decreases to $1$~mCrab for a total exposure of $1$~Ms.  
\textit{Systematic uncertainties:}
\begin{itemize}
\item 0.16$\%$ ($4.0$~mCrab) -- root-mean-squared residuals after
  background model subtraction from the observed count rate
  (Fig.~\ref{fig:uncert}),
\item 0.09$\%$ ($2.3$~mCrab) -- CXB variance per IBIS FOV,
\item 0.16$\%$ ($4.0$~mCrab) -- CXB variance per GA survey area.
\end{itemize}
The systematics quadratically add to the error of the measurement. The
total systematic error owing to the background model and CXB variance (per GA 
survey area) is $5.7$~mCrab.

\end{appendix}

\end{document}